# Multiplicative noise and charge density wave dynamics in $Rb_{0.30}MoO_3$

J. Dumas, J. Marcus,

Institut Néel, CNRS and Université Joseph Fourier, BP 166, 38042 Grenoble Cedex 9, France

**Abstract**

We have investigated the combined effects of a multiplicative noise and nonlinear conductivity on the charge density wave (CDW) dynamics of the quasi-one dimensional conductor $Rb_{0.30}MoO_3$. When the amplitude of a bipolar rectangular voltage pulse excitation is modulated by a multiplicative Gaussian white noise perturbation, the average pulse amplitude is smaller than the unperturbed one and shows a non monotonous behaviour when the pulse amplitude is increased. We analyze the transient response in terms of an increase of phase coherence of the charge-density-wave.

**Introduction**

Noise in some nonlinear systems can play a counterintuitive role. Noise can lead to ordered states through its interactions with nonlinearities of the system. Typical examples are stochastic resonance (SR) or noise induced phase transitions [1, 2]. A quasi-one dimensional CDW conductor is an interesting non linear system where such phenomena can be explored.

At low electric field, the CDW is prevented from moving by pinning to randomly distributed impurities. Above a depinning threshold $E_t$, the CDW starts to slide. The sliding of the CDW results in nonlinear conductivity accompanied by generation of a broad band noise (BBN) with spectral density $S(f) \sim f^{-\alpha}$ ($\alpha \sim 1$) and high frequency coherent oscillations, the so-called narrow-band noise (NBN) [3].

In a previous work [4], we have shown that SR effects may be involved in the depinning process of a charge density wave (CDW) in the quasi-one dimensional conductor $K_{0.30}MoO_3$ in the presence of an external additive white noise superimposed on a dc bias voltage. In a recent work [5], we have studied in the same conductors the transient CDW voltage response $V(t)$ to bipolar symmetrical rectangular current pulses when the pulse amplitude is modulated by a white noise with a modulation depth ranging from 20% up to 80 %. When the multiplicative white noise perturbation is combined with the transient voltage pulse excitation, the average pulse amplitude response is found to be smaller than the unperturbed one and shows a non monotoneous behaviour when the pulse amplitude is increased. Our results were discussed in terms of SR, a phenomenon observed in a large variety of nonlinear systems with a fluctuating environment. We have now performed similar complementary measurements on $Rb_{0.30}MoO_3$ samples. The results viewed as multiplicative stochastic processes are discussed in terms of increased phase coherence of the CDW.

**Results**

The experimental arrangement consists of Tektronix TDS 5034 B Digital Phosphor Oscilloscope and Agilent 33120 A function generator. Rectangular bipolar periodic voltage pulses were applied to the sample at T=77K using a four-probe geometry. Large resistance was applied in series with the pulse generator to achieve a current driven arrangement. The experiments were performed with a pulse width $\tau$=1 ms.

The pulse amplitude is modulated using a Gaussian white noise (bandwidth 10 MHz) with a modulation depth ranging from 20% up to 80 %. The modulation depth is defined as the ratio of the peak-to-peak value of the modulation signal amplitude to the pulse amplitude $V_{pp}$. The multiplicative noise amplitude increases linearly with the excitation V(t). The pulse response was averaged over 2048 scans.

We show in Figure 1 the ratio $\Delta V/V = [V_{pp}(D=0)-V_{pp}(D)] /V_{pp}(D=0)$ as a function of $V_{pp}(D=0)$ for a $K_{0.30}MoO_3$ sample and for noise modulation depths D=20%, 40% and 60%. $V_{pp}(D)$ is the voltage response obtained after averaging over 2048 steps. For given noise amplitudes characterized by the modulation depths D, $V_{pp}(D)$ is found to be always smaller than $V_{pp}(D=0)$. A non monotonous behavior of $\Delta V/V$ with $V_{pp}(D=0)$ is observed. $\Delta V/V$ starts to increase just above the threshold voltage, then shows a broad maximum above $V_t$. A small shift in the maxima of $\Delta V/V$ can be observed, as the noise D is increased.

The nonlinear characteristics V(I) for two $Rb_{0.30}MoO_3$ samples were obtained by dc measurements. The threshold voltage was $V_t$=12 mV at 77K corresponding to a threshold field $E_t$ = 120 mV/cm for both samples. The normalized conductivity $\sigma/\sigma_0$, with $\sigma$= I/V and $\sigma_0$ the conductivity in the Ohmic regime, is plotted in Figure 2 as a function of the normalized voltage $V/V_t$ for samples #1 and #2. These results indicate that the excess CDW current is smaller in sample #2 than in sample #1.

Figure 3(a) shows the relative change of the ratio $\Delta V/V$ as a function of $V_{pp}(D=0)$ for D=40% and 80%.in $Rb_{0.30}MoO_3$ (sample #1). The increase in $\Delta V/V$ just above threshold ($V_t$=12 mV) is sharper in this $Rb_{0.30}MoO_3$ sample than in $K_{0.30}MoO_3$. $\Delta V/V$ shows a well defined maximum then decreases slowly. A gradual increase of $\Delta V/V$ at threshold is found in $Rb_{0.30}MoO_3$ (sample #2), as shown in Figure 3(b). These results corroborate the previous ones obtained on $K_{0.30}MoO_3$.

**Discussion**

While the internal broad band $1/f^\alpha$ noise generated above threshold can be viewed as an additive noise source, the external applied Gaussian white noise is a multiplicative noise since its amplitude D is proportional to the pulse amplitude. D represents a fluctuating force between V(1+D) and V(1-D) with $<D(t)> = 0$ and D(t) uncorrelated in time. The system is driven simultaneously by noise and a deterministic transient force V(t). Additive noise can never be switched off at given voltage while for an external noise it is possible.

The increase of $\Delta V/V$ with $V_{pp}$(D=0) indicates that the excess CDW current increases weakly when the external noise is turned on. We propose that noise is homogeneizing the CDW phase throughout the sample. This noise enhanced phase coherence is reminiscent of stochastic resonance [1]. In nonlinear systems, this counterintuitive phenomenon arises from an interplay between noise and a deterministic external periodic signal, the repetitive pulse excitation in our experiments.

Numerical simulations have been reported for the effects of an additive noise on the CDW dynamics in the context of the Fukuyama-Lee-Rice (FLR) model for deformable CDW's or of the rigid model and using a Langevin equation for the overdamped CDW phase dynamics:

$d\phi(t)/dt = - d\mathcal{H}/d\phi + \Gamma(t)$ where $\mathcal{H}$ is the FLR Hamiltonian :

$$\mathcal{H} = \int dx [\frac{K}{2}(\frac{d\phi}{dx})^2 - \sum_i V_i \delta(x-R_i)\rho.\cos(2k_F x + \phi(x)) + Ex\frac{d\phi}{dx}]$$

K is the stiffness constant of the CDW, $2k_F$ the wavevector modulation, $V_i$ the pinning potential of the CDW with impurity at position $R_i$ and E the electric field.

$\Gamma(t)$ is the noise with $<\Gamma(t)> = 0$ and is delta-correlated in time t, $<\Gamma(t)\Gamma(t')>=D\delta(t-t')$. The force is $f = - d\mathcal{H}/d\phi$.

The associated time dependent ordinary Fokker-Planck equation for the probability density $P(\phi,t)$ of the stochastic variable $\phi$ can be written:

$\delta_t P(\phi,t) = - \delta_\phi[fP(\phi,t)] + D\delta^2_{\phi\phi}P(\phi,t)$

Within the FKL model for 1D discretized lattice of overdamped deformable CDW's [6] a random force for heat bath is written $<\xi_i(t)> = 0$  $<\xi_i(t)\xi_j(t') = 2\pi\gamma^2 T\delta_{ij}\delta(t-t')$. The authors find an increase of the normalized conductivity $\sigma/\sigma_0$ and a decrease of $E_t$ as the noise effective temperature T increases. A decrease of $E_t$ and rounding of the depinning transition with noise has been reported by Lee [7]. Konno [8] has reported a strengthening of spatial phase coherence under the effect of additive Gaussian white noise. Sweetland et al. [9] have considered a Langevin equation to describe fluctuations of X-ray scattering intensities.

A current noise has been considered in the study of depinning where the noise $T_N$ has an amplitude proportional to the CDW current amplitude $T_N \alpha \; <\overline{d\varphi/dt}>$ where the bar denotes a time average [10-13].

The effect of fluctuations in the context of a rigid CDW has been investigated for a thermal noise [14-17] and for a current noise [18-20]. For a Gaussian white noise, Reguera et al. [21] have found an increase of the average CDW velocity with noise.

The constructive ordering role of a multiplicative noise can be discussed also in terms of the first order stochastic differential Langevin equation. However, a multiplicative noise situation is a more complex situation than that described by Langevin equations with additive noise. In the case of a system driven by a multiplicative Gaussian white noise the Langevin equation requires an appropriate interpretation of the integral of the noise term [22-25]. The parameter that determines the points of time at which $\Gamma$ is evaluated in the integral has to be specified. One encounters the so-called Ito-Stratonovich dilemma when choosing the appropriate calculus for integrating a multiplicative Langevin equation. Different time discretization methods lead to different results. For systems with additive noise the Ito–Stratonovich dilemma does not exist; both approaches should yield the same results.

In the Langevin equation with multiplicative noise :

$d\phi(t)/dt = f(\phi(t)) + g(\phi(t))\Gamma(t)$ (Eq. 1)

the strength prefactor of the noise term $\Gamma(t)$ depends on the state variable $\phi$, f= -dU($\phi$,t)/d$\phi$ is the external force, U is a deterministic potential. f, g are deterministic functions and $\Gamma$ is a Gaussian white noise with zero average <$\Gamma$(t)> =0 and covariance <$\Gamma$(t)$\Gamma$(t')> = 2D$\delta$(t-t'). <….> indicates an average over the noise distribution.

For an additive noise g=1 while for multiplicative noise g is not constant. The effect of noise depends on the state of the system.

Due to the noise, for each jump in the solution $\phi$ of the Langevin equation, the value of $\phi$ and hence of g($\phi$) is undetermined. In any time interval, fluctuations occur an infinite number of times. We have to choose how to interpret the product g($\phi$)$\Gamma$(t) and the Fokker-Planck equation will depend on that interpretation and will give different results. One possibility is to choose the value of g(ϕ) either just before the jump, after the jump or the mean value. Taking the mean value corresponds to the Stratonovich scenario while taking the value just before the jump is the Ito scenario.

For an additive noise the corresponding drift-diffusion Fokker-Planck equation for the temporal evolution of the probability density P(ϕ,t) of the variable ϕ associated with the Langevin equation :

$d\phi(t)/dt = f(\phi(t)) + \Gamma(t)$

can be written:

$\delta_t P(\phi,t) = -\delta_\phi[f(\phi,t)P] + D\delta^2_{\phi\phi}P(\phi,t)$

with a drift term $-\delta_\phi[f(\phi,t)P]$ and a diffusion term $D\delta^2_{\phi\phi}P(\phi,t)$.

For multiplicative noise (Eq.1) in the Ito scenario:

$\delta_t(P(\phi,t) = -\delta_\phi[f(\phi,t)P] + D\delta^2_{\phi\phi}[g^2(\phi,t)P]$ with $<\Gamma(t)\Gamma(t')> = 2D\delta(t-t')$

and in the Stratonovich scenario :

$\delta_t(P(\phi,t) = -\delta_\phi[f(\phi,t)P] + D\delta_\phi[g\delta_\phi(gP)]$

The two equations differ by a noise indued additional "spurious drift term". The Ito and Stratonovich interpretations coincide when g = constante.

In this context, the shift of the maxima of ΔV/V when the noise D is increased (see Fig.1 and Fig. 6(a) in ref. [5]) would correspond to a shift in the maxima of the probability distribution. One should note that a drift and diffusion of a moving CDW in NbSe$_3$ has been reported by Gill [26].

**Conclusion**

In summary, we have experimentally investigated nonlinear effects of a multiplicative external noise on the CDW driven by a rectangular bipolar signal in Rb$_{0.30}$MoO$_3$. We have corroborated previous results on K$_{0.30}$MoO$_3$. We suggest that the non monotonous behaviour of ΔV/V with V$_{pp}$(D=0) for a given noise modulation depth D is consistent with the stochastic resonance phenomenon. Theoretical studies based on Fokker Planck equation in the Ito or Stratonovich description of the effect of a multiplicative noise on the CDW phase coherence would be useful.

**Figure 1.** [$V_{pp}(D=0)-V_{pp}(D)$] /$V_{pp}(D=0)$ as a function of $V_{pp}(D=0)$ for noise amplitude D = 20, 40, 60% for a $K_{0.30}MoO_3$ sample. T=77K. $V_t$=20 mV. From ref. [5].

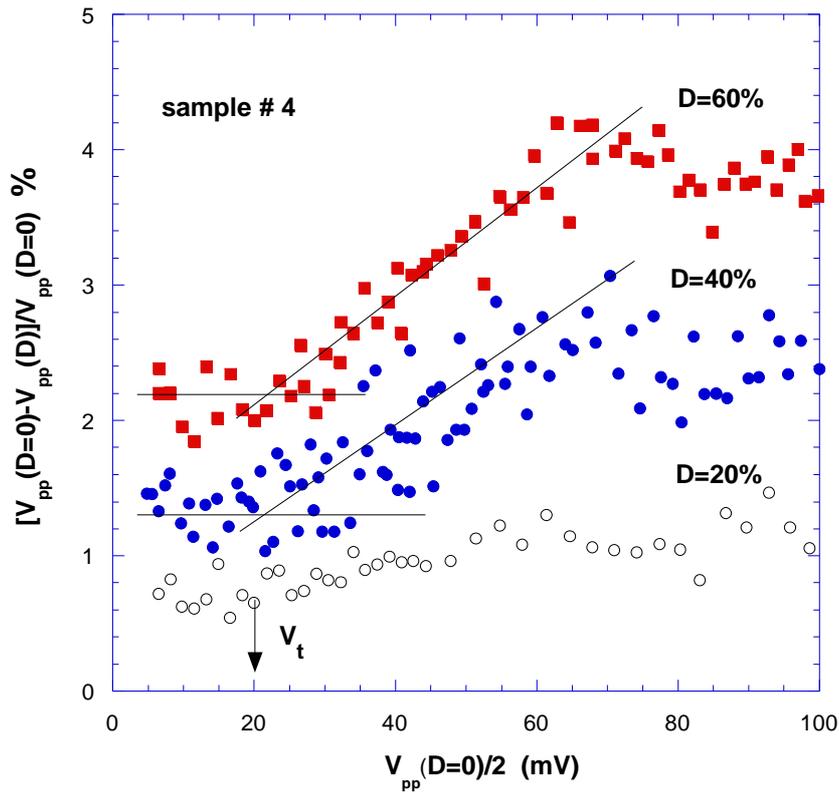

**Figure 2**. Normalized conductivity $\sigma/\sigma_0$ as a function of normalized voltage $V/V_t$ for $Rb_{0.30}MoO_3$ samples #1 and #2. T=77K. Threshold voltages: $V_t = 12$ mV.

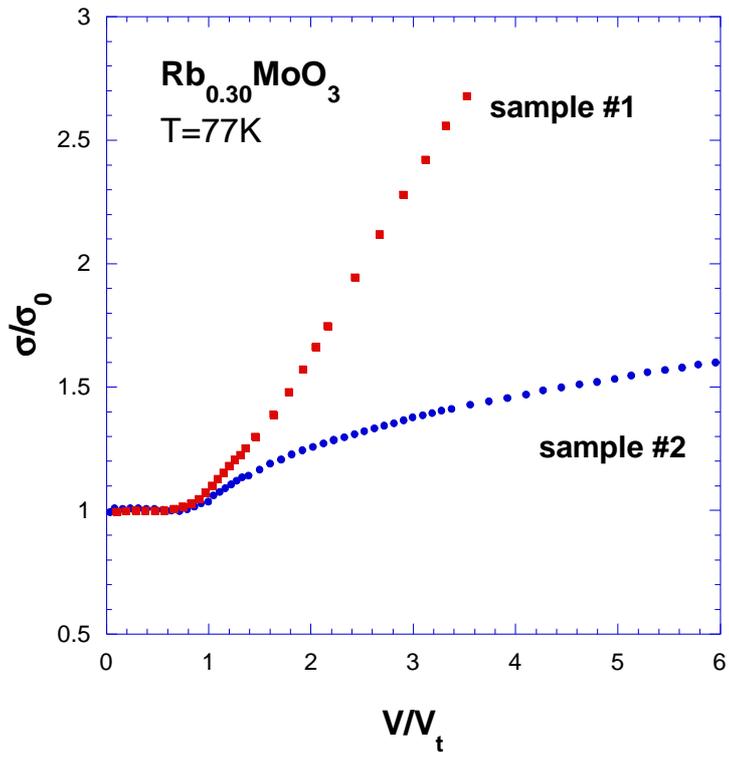

**Figure 3 (a).** [$V_{pp}(D=0)-V_{pp}(D)$] /$V_{pp}(D=0)$ as a function of $V_{pp}(D=0)$ for noise amplitude D=40% and D=80% for $Rb_{0.30}MoO_3$ (sample #1). $V_t$ = 12 mV. T=77 K.

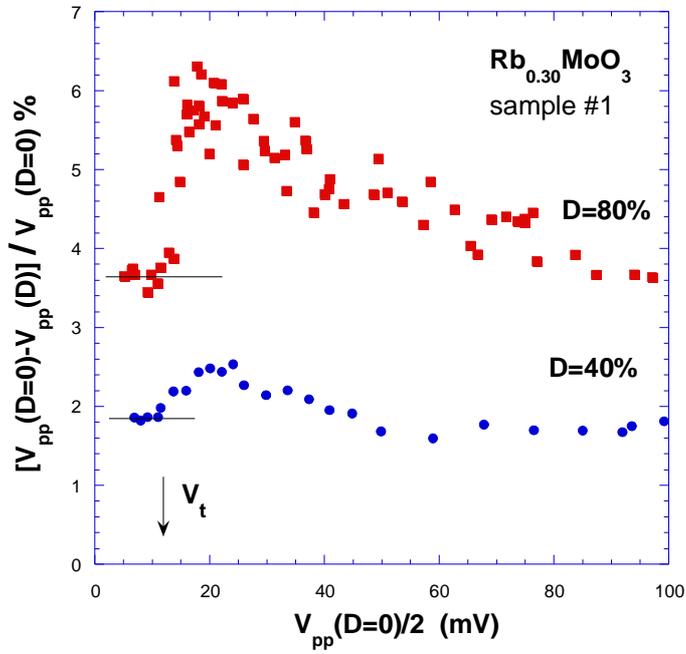

**Figure 3 (b).** Same as in Figure 3(a) for $Rb_{0.30}MoO_3$ (sample #2) and D=40%. $V_t$ = 12 mV. T=77 K.

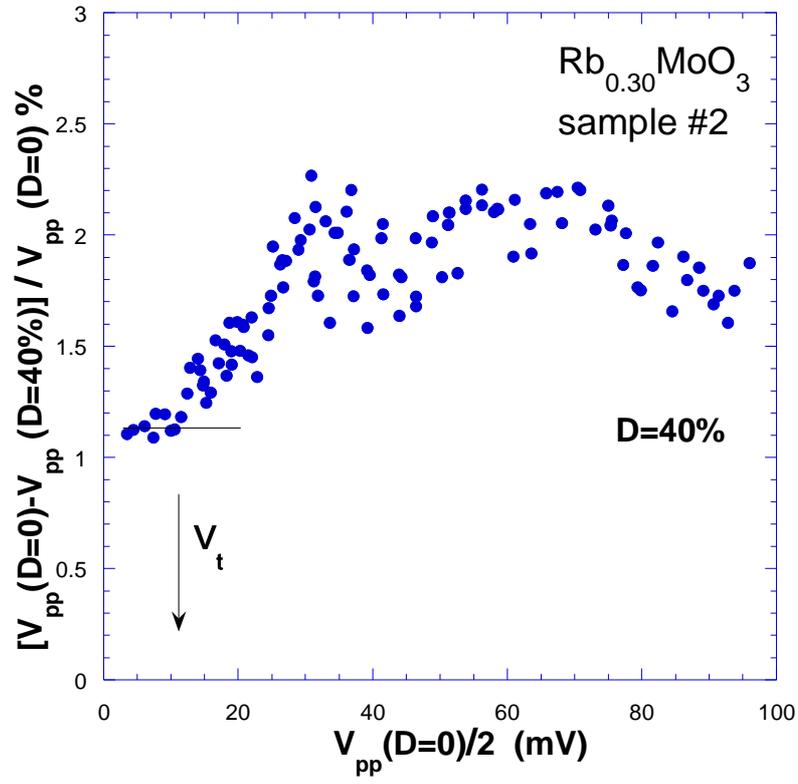